\overfullrule=0pt
\input harvmac
\def\a{{\alpha}}
\def\ad{{\dot \alpha}}

\def\L{{\Lambda}}
\def\b{{\beta}}
\def\g{{\gamma}}
\def\d{{\delta}}
\def\cG{{\cal G}}

\def\s{{\sigma}}

\def\O{{\Omega}}
\def\half{{1\over 2}}
\def\th{{1\over 3}}
\def\tth{{2\over 3}}
\def\p{{\partial}}
\def\t{{\theta}}
\def\tb{{\overline\theta}}

\Title{\vbox{\hbox{IFT-P.060/2001 }}}
{\vbox{
\centerline{\bf The Ramond Sector of Open Superstring Field Theory}}}
\bigskip\centerline{Nathan Berkovits\foot{e-mail: nberkovi@ift.unesp.br}}
\bigskip
\centerline{\it Instituto de F\'\i sica Te\'orica, Universidade Estadual
Paulista}
\centerline{\it Rua Pamplona 145, 01405-900, S\~ao Paulo, SP, Brasil}

\vskip .3in
Although the equations of motion for the Neveu-Schwarz (NS) and
Ramond (R) sectors of open superstring field theory can be covariantly
expressed in terms of one NS and one R string field, picture-changing 
problems prevent the construction of an action involving these two string 
fields. However, a consistent action can be constructed by dividing the NS 
and R states into three string fields which are real, chiral and antichiral. 

The open superstring field theory action includes a WZW-like term for the 
real field and holomorphic Chern-Simons-like terms for the chiral and 
antichiral fields. Different versions of the action can be constructed 
with either manifest d=8 Lorentz covariance or manifest N=1 d=4 
super-Poincar\'e covariance. The lack of a manifestly d=10 Lorentz 
covariant action is related to the self-dual five-form in the Type IIB 
R-R sector.

\vskip .3in

\Date {September 2001}

\newsec{Introduction}

Open superstring field theory has recently returned to center stage
due to the work of Sen and others on tachyon condensation \ref\Sen
{A. Sen, {\it Universality of the Tachyon Potential},
JHEP 9912 (1999) 027, hep-th/9911116.}. Although
the Ramond sector of open superstring field theory is not directly related
to tachyon condensation, an understanding of this sector is crucial
for studying other properties of the superstring action. For example,
it will be argued below that certain terms involving the Ramond sector
are expected to satisfy non-renormalization theorems.

Although a cubic open superstring field theory action involving both the
Neveu-Schwarz (NS) and Ramond (R) sectors was proposed in 
\ref\witten{E. Witten,
{\it Interacting Field Theory of Open Superstrings},
Nucl. Phys. B276 (1986) 291.}, this action (as well as all others
\ref\others{S. Mandelstam, {\it Interacting String Picture of the
Neveu-Schwarz-Ramond Model}, Nucl. Phys. B69 (1974) 77\semi
C.R. Preitschopf, C.B. Thorn and S.A. Yost, {\it Superstring Field
Theory}, Nucl. Phys. B337 (363) 1990\semi N. Berkovits, M.T. Hatsuda
and W. Siegel, {\it The Big Picture},
Nucl. Phys. B371 (1992) 434, hep-th/9108021\semi
I.Ya. Arefeva, D.M. Belov, A.S. Koshelev and P.B. Medvedev,
{\it Tachyon Condensation in Cubic Superstring Field Theory},
hep-th/0011117.}
constructed using picture-changing operators) suffers from problems
caused by the presence of picture-raising and picture-lowering operators.
The picture-raising operator appears in the NS interaction term and
leads to contact-term divergences in tree-level scattering amplitudes
\ref\wendt
{C. Wendt, {\it Scattering Amplitudes and Contact Interactions in
Witten's Superstring Field Theory}, Nucl. Phys. B314 (1989) 209\semi
J. Greensite and F.R. Klinkhamer, {\it New Interactions for Superstrings},
Nucl. Phys. B281 (1987) 269.}. The picture-lowering operator appears
in the R kinetic term and, as will be discussed in section 2, 
leads to a breakdown
of gauge invariance due to its nontrivial kernel. Another problem
with the picture-lowering operator is that it does not commute with the
$b$ ghost and therefore cannot appear in the closed superstring kinetic
term for the R-NS, NS-R or R-R sectors. 

As discussed in earlier papers \ref\sft{
N. Berkovits, {\it Super-Poincare Invariant Superstring Field Theory},
Nucl. Phys. B450 (1995) 90, hep-th/9503099.}
\ref\review{N. Berkovits, {\it
A New Approach to Superstring Field Theory},
Fortschritte der Physik (Progress of
Physics) 48 (2000) 31, hep-th/9912121\semi
N. Berkovits and C.T. Echevarria, {\it
Four-Point Amplitude from Open Superstring
Field Theory}, Phys. Lett.   
B478 (2000) 343, hep-th/9912120\semi N. Berkovits, {\it
Review of Open Superstring Field Theory}, hep-th/0105230.},
the above picture-changing problems can be avoided by
working in the large RNS Hilbert space \ref\fms{D.  
Friedan, E. Martinec and S. Shenker, {\it
Conformal Invariance, Supersymmetry and String Theory},
Nucl.  Phys B271 (1986) 93.} which includes the $\xi$ zero mode
coming from fermionizing the $(\beta,\g)$ ghosts.
The NS contribution to the
field theory action resembles a Wess-Zumino-Witten
model where the group generator
$g$ is related to the NS string field $\Phi$ by $g=e^\Phi$ and multiplication
of string fields in the exponential uses Witten's midpoint interaction
\ref\midp{E. Witten, {\it Noncommutative Geometry and String Field Theory},
Nucl. Phys. B268 (1986) 253.}. 

A natural question is how to include the Ramond contribution to
the superstring field theory action. This question was partially
answered in \sft\ where a manifestly
N=1 d=4 super-Poincar\'e covariant action was constructed by splitting
the NS and R states into
a real, chiral and antichiral string field. However, the resulting
action was quite complicated and it was unclear if other actions
could be constructed which preserve more symmetries.

In this paper, it will be argued that splitting the NS and R states
into three string fields is necessary for constructing consistent 
superstring field theory actions.
Although the open superstring field theory equations
of motion can be covariantly expressed in terms of a single NS and R
string field, it is not possible to construct a consistent action 
out of these two string fields \foot{In fact, 
the superstring field theory equations of
motion can also be expressed
in terms of a single string superfield in a manifestly d=10
super-Poincar\'e covariant manner \ref\pure{N. Berkovits, {\it
Super-Poincar\'e Covariant Quantization of the Superstring}, JHEP 04 
(2000) 018, hep-th/0001035.}. However, it does not appear possible
to construct an action in terms of this single string superfield.}. 
But
by using three string fields consisting of a real, chiral and
antichiral string field, a consistent action can be constructed which
reproduces the desired equations of motion. 
This action includes a WZW-like
term constructed from the real string field, a kinetic term 
for the chiral and antichiral string fields coupled minimally
to the real field, and a holomorphic and antiholomorphic
Chern-Simons-like term 
constructed from the chiral and antichiral fields. 
The holomorphic Chern-Simons-like term involves integration over
a chiral subspace and therefore resembles a superspace F-term.
For the usual reasons, this F-term is expected to satisfy non-renormalization
theorems.

Depending on how
the NS and R states are distributed among the three string fields, 
different subgroups of d=10 super-Poincar\'e covariance can
be manifestly preserved. For example, in a flat background, superstring
field theory actions can be constructed which manifestly preserve 
either d=8 Lorentz covariance or N=1 d=4 super-Poincar\'e covariance.
The difficulty in constructing actions with manifest d=$4k+2$ Lorentz
covariance is related to the presence of self-dual $(2k+1)$-forms in
d=$4k+2$.

In section 2 of this paper, 
equations of motion will be defined 
using a single NS and R string field, and it will be argued
that they cannot come from varying an action.
In section 3,
an open superstring field theory action will be constructed by splitting
the superstring states into three string fields. 
In section 4, it will be shown how different choices for
splitting the superstring states into three string fields produce
actions with different manifest symmetries.
And in section 5, some open questions will be discussed including
the construction of a closed superstring field theory action.

\newsec{Problems using Two String Fields}

\subsec{Equations of motion}

Although it will not be possible to construct a consistent action
using a single NS and R string field, one can define equations of
motion using these two string fields. 
In the action of \witten, the ghost-number one
\foot{Unlike \fms\ where $j_{ghost}=cb+\p\phi$, the ghost number current
will
be defined here as $j_{ghost}=cb+\eta\xi$. So $(\eta,\xi)$ carry
ghost number $(+1,-1)$ and picture $(-1,+1)$ while
$e^{n\phi}$ carries ghost
number zero and picture $n$. This definition of ghost number
agrees with that of \fms\
at zero picture, but has the advantage of commuting with picture-changing
and spacetime supersymmetry.} NS and R string fields are defined
in the small RNS Hilbert space and carry picture $-1$ and $-\half$.
For example, 
the massless gluon $A_m(x)$ and gluino $\chi^\a(x)$
are represented by the
vertex operators $c e^{-\phi} \psi^m A_m(x)$ and
$c e^{-\half \phi} \Sigma_\a\chi^\a(x)$ where $\Sigma_\a$
is the RNS spin field.  However, as was shown in \sft, it is more
convenient for constructing actions to use string fields in the
large RNS Hilbert space at ghost number zero. To linearized level,
these ghost number zero string fields are related to the ghost
number one string fields of \witten\ by adding the $\xi$ zero mode.
Since $\xi$ carries picture $+1$, it is natural to define the
NS and R string fields, $\Phi$ and $\Psi$, to carry picture 0 and $\half$. 
In other words,
the massless gluon and gluino will be
represented by the
vertex operators $\Phi = \xi c e^{-\phi} \psi^m A_m(x)$ and
$\Psi = \xi c e^{-\half \phi} \Sigma_\a\chi^\a(x).$ 

As discussed in 
\ref\topo{N. Berkovits and C. Vafa, {\it N=4 Topological Strings},
Nucl. Phys. B433 (1995) 123, hep-th/9407190.},
the linearized
equations of motion and gauge invariances for ghost number zero string
fields in the large Hilbert space can be written 
\eqn\lineq{Q\widetilde\eta \Phi = 
Q\widetilde\eta \Psi = 0, \quad
\d\Phi = Q\Lambda_0 +\widetilde\eta\L_1, \quad \d\Psi=Q\L_{\half}+
\widetilde
\eta \L_{ {3\over 2}},}
where $Q=\int[c(T_m-b\p c -\p^2\phi -\half(\p\phi)^2 +\eta\p\xi)
+\eta e^\phi G_m -\eta\p\eta e^{2\phi} b]$ 
is the RNS BRST operator, $\widetilde\eta$ denotes the zero mode of
the $\eta$ ghost\foot{To avoid confusion with the picture subscript,
the notation $\widetilde\eta$ will be used instead of $\eta_0$.}, 
and $\L_n$ are ghost-number $-1$
string fields of picture $n$.
Note that the linearized
gauge transformations parameterized by $\L_1$ and $\L_{3\over 2}$
are necessary if one uses string fields in the large Hilbert space.
One would now like to find a nonlinear version of \lineq\ which does
not involve picture-changing operators. One can check that
equation \lineq\ generalizes to the following nonlinear equations
of motion and gauge invariances:
\eqn\nonlineq{\widetilde
\eta (e^{-\Phi} (Q e^\Phi)) = -(\widetilde
\eta\Psi)^2,\quad
Q(e^{\Phi}(\widetilde
\eta\Psi) e^{-\Phi}) = 0,}
$$\d e^\Phi = e^\Phi (\widetilde
\eta\L_1 - \{\widetilde
\eta\Psi, \L_\half\}) +(Q\L_0)e^\Phi,\quad
\d\Psi = \widetilde
\eta\L_{3\over 2} + [\Psi,\widetilde
\eta\L_1] + \{Q + e^{-\Phi}(Q e^\Phi),
\L_\half\}. $$

Although these equations of motion and gauge invariances appear complicated,
they simplify when expressed in terms of the ghost number one string
fields $V = e^{-\Phi} (Q e^\Phi)$ and $\O= \widetilde
\eta\Psi$. In terms of $V$ and
$\O$, \nonlineq\ implies
\eqn\cseq{QV = - V^2, 
\quad Q\O = -\{\O,V\},
\quad \widetilde
\eta V = -\O^2, \quad \widetilde
\eta \O=0,}
$$\d V = [Q +V, \widetilde
\eta\L_1 - \{\O,\L_\half\} ],
\quad \d\O = \widetilde
\eta(  \{Q+V,\L_\half\} - \{\O,\L_1\}).$$
Note that the linearized contributions to \cseq\ are the standard
BRST conditions for ghost number one vertex operators in the
small Hilbert space, i.e.
\eqn\linstand{QV= \widetilde
\eta V= Q\O = \widetilde
\eta\O =0,\quad
\d V = Q\widetilde
\eta\L_1,\quad \d \O = \widetilde
\eta Q\L_\half.}

The above equations of motion and gauge invariances can be simplified
even further by defining 
\eqn\simpl{G= G_0 + G_{-1}= Q + \widetilde
\eta \quad{\rm and}
\quad A = A_0 + A_{-\half} = V+ \O = e^{-\Phi}(Q e^\Phi) + \widetilde
\eta\Psi} 
where $G_n$ and $A_n$ carry picture $n$.
Then \nonlineq\ and \cseq\ are equivalent to
\eqn\coveq{ (G +A)^2=0,\quad \d A = G\s + [A,\s],}
where $\s= \s_0 + \s_\half=
\widetilde
\eta\L_1 + \{Q+V,\L_\half\}$. By expanding \coveq\ into
the different pictures which contribute, 
one recovers equations \nonlineq\ and \cseq.
In other words, the four equations in the first line of \cseq\
are implied by the contributions to
$(G+A)^2=0$ at picture $[0,-\half,-1,-{3\over 2}]$.
The gauge transformations parameterized by $\L_\half$ and $\L_1$
transform $A$ as in \coveq\ while those parameterized by
$\L_0$ and $\L_{3\over 2}$ leave $A$ invariant.

\subsec{Problems with an action}

Although the equations of \coveq\ are an obvious analog of the bosonic string
field theory
equations of motion and gauge invariances, picture changing problems
prevent the construction of an action which yields these equations of
motion. If one sets to zero the Ramond string field, a Wess-Zumino-Witten-like
action can be constructed out of the NS string field $\Phi$ which
yields the desired equation of motion $\widetilde
\eta(e^{-\Phi}(Q e^\Phi))=0.$
However, there is no way to consistently include a single Ramond string
field $\Psi$ into this action.

This impossibility can already be seen by analyzing the kinetic term for
the Ramond string field. To recover the linearized equation of motion
$Q\widetilde
\eta\Psi=0$, one would need a kinetic term $\langle \Psi Q\widetilde
\eta \Psi\rangle$.
However, since the large Hilbert space norm is defined as
$\langle \xi e^{-2\phi} c\p c \p^2 c \rangle =1$, a non-vanishing kinetic 
term must carry picture $-1$. But since $\Psi$ cannot
carry zero picture, $\langle\Psi Q\widetilde
\eta\Psi\rangle$ cannot carry picture $-1$.

One possible solution \ref\yam{J. Yamron, {\it A Gauge Invariant Action
for the Free Ramond String}, Phys. Lett. B174 (1986) 69.}\witten\
would be to insert the picture-lowering operator $Y= c\p\xi e^{-2\phi}$
at the midpoint of the
kinetic term as 
\eqn\openrk{\langle\Psi Y(\pi) Q \widetilde
\eta\Psi\rangle.} 
However, this
would produce the equation of motion $Y(\pi)Q\widetilde
\eta\Psi=0$ and, since $Y(\pi)$
has a nontrivial kernel,
would not imply $Q\widetilde
\eta\Psi=0$.
Note that the kernel of 
$Y(\pi)$ only includes string states which are singular at
their midpoint and therefore does not include ``smooth'' states,
i.e. states constructed from the ground state with a finite number
of mode operators. However, since the product of two smooth states
is not necessarily smooth, the kernel of $Y(\pi)$ is nontrivial in
the complete Fock space of string states.

One could try
to truncate out all string states in the kernel of $Y(\pi)$ \ref\aref
{I.Ya. Arefeva and P.B. Medvedev, 
{\it Anomalies in Witten's Field Theory of the NSR String},
Phys. Lett. B212 (1988) 299\semi
I.Ya. Arefeva and P.B. Medvedev,
{\it Truncation, Picture Changing Operation and Spacetime Supersymmetry
in Neveu-Schwarz-Ramond String Field Theory}, 
Phys. Lett. B202 (1988) 510.},
but such a truncation would ruin the
associativity properties of the midpoint interaction, leading to
a breakdown of gauge invariance. In other words,
if the operator $T$ truncates out states in the kernel of $Y(\pi)$,
the truncated product of three smooth string states $[A,B,C]$ is
either $T(T(A\times B)\times C)$ or 
$T(A\times T(B\times C))$, which depends on the order of multiplication.
One could also try to gauge away all states in the kernel of $Y(\pi)$
since the kinetic term of \openrk\ is invariant under $\d\Psi=\L$
for $\L\in\ker Y(\pi)$. However, there is no such gauge transformation
which also leaves invariant the interaction terms.
So just as
insertion of the
picture-raising operator $X=\{Q,\xi\}$ creates
inconsistencies due to
contact term divergences, insertion of
the picture-lowering operator $Y=c\p\xi e^{-2\phi}$
creates inconsistencies due to its nontrivial kernel.

Furthermore, the natural generalization of the
R kinetic term of \openrk\ to closed
superstring field theory is the R-R kinetic term
\eqn\rrk{\langle\Phi_{RR} Y\widehat Y (c-\widehat c)_0 (Q +\widehat Q)
\widetilde
\eta\widetilde{\widehat\eta}~\Phi_{RR}\rangle}
where the unhatted and hatted operators are left and right-moving,
and $\Phi_{RR}$ is the R-R closed string field at ghost
number zero and (left,right)-moving
picture $(\half,\half)$.  However, this
kinetic term is not gauge invariant since $Y\widehat Y$ does not
commute with $(b-\widehat b)_0$. Recall 
that closed string fields
$\Phi$ must satisfy the constraint $(b-\widehat b)_0\Phi=0$
and the gauge transformation $\d\Phi=(Q+\widehat Q)(b-\widehat b)_0\Lambda$
only leaves the action invariant if $\{(b-\widehat b)_0, 
[Q+\widehat Q,{\cal O}]\}=0$ where ${\cal O}$ is the kinetic operator
\ref\zw{
B. Zwiebach, {\it Closed String Field Theory: Quantum Action and
the Batalin-Vilkovisky Master Equation}, Nucl. Phys. B390 (1993)
33, hep-th/9206084.}. 
So \rrk\ is inconsistent even before worrying about the nontrivial kernel
of $Y\widehat Y$.

An alternative method for constructing a Ramond kinetic term 
is to split the Ramond states into
two string fields, $\Psi$ and $\overline\Psi$, where $\Psi$ is defined
to carry picture $+\half$ and $\overline
\Psi$ is defined to carry picture $-\half$. 
Although this method necessarily breaks manifest d=10 Lorentz
covariance (since the sixteen component d=10 spinor is broken into
two eight component spinors), 
it allows one to construct the non-vanishing kinetic term
$\langle\Psi Q\widetilde
\eta\overline
\Psi\rangle$. As will be seen in the following
sections, such a solution beautifully generalizes to a full nonlinear 
open superstring field theory action.

\newsec{Open Superstring Formalism with Three String Fields}

\subsec{Equations of Motion}

Since one now has three string fields $[\Phi,\Psi,\overline
\Psi]$, one has
to decide how the NS and R states are distributed among these fields.
Suppose that
one can define a conserved charge
$C$ such that all superstring
states carry one of three distinct values of this
charge. Normalizing $C$ such that $\eta$ carries charge $C=-1$,
it will be argued below that one can only construct consistent actions
if these three distinct $C$-charges are $C=0$, $C={1\over 3}$ 
and $C=-{1\over 3}$.
States with charge $C=0$ will be
represented by $\Phi$, states with charge $C={1\over 3}$ 
will be represented by
$\Psi$, and states with charge $C=-{1\over 3}$ 
will be represented by $\overline
\Psi$. The hermiticity properties of these three string fields will
be discussed at the end of subsection (3.3). 

So just as picture was used in the previous section
to distribute states among two string fields, $C$-charge will be used here
to distribute states among three string fields.
As will be shown in section 4, different choices for $C$ 
produce different actions with different manifest symmetries.
Although there will be no d=10 Lorentz invariant choice
of $C$, there are choices which preserve either d=8 Lorentz invariance or
N=1 d=4 super-Poincar\'e invariance.

To construct an action in terms of the three string fields 
$[\Phi,\Psi,\overline
\Psi]$, one first needs to find nonlinear equations of
motion and gauge invariances which generalize the linearized equations
\eqn\tlin{
Q\widetilde
\eta \Phi = 
Q\widetilde
\eta \Psi = Q\widetilde
\eta\overline
\Psi= 0, }
$$
\d\Phi = Q\Lambda_0 +\widetilde
\eta\L_1, \quad \d\Psi=Q\L_{1\over 3}+
\widetilde
\eta \L_{4\over 3}, \quad
\d\overline
\Psi = Q\L_{- {1\over 3}}+
\widetilde
\eta \L_{2\over 3} $$
where $\L_n$ carries $C$-charge $n$. To simplify the discussion, $Q$
has temporarily been assumed to carry zero $C$-charge although this
assumption will later be relaxed in subsection (3.3). 

Following the discussion of the previous section, one would
like to define a gauge field $A$ in terms of
$[\Phi,\Psi,\overline
\Psi]$ such that $(G+A)^2=0$ gives the nonlinear
equations of motion. Defining
\eqn\deft{G=G_0+G_{-1}=Q+\widetilde
\eta,\quad A=A_0+A_{-{1\over 3}}+A_{-{2\over 3}}
= e^{-\Phi}(Q e^\Phi)+ e^{-\Phi}(Q\overline
\Psi)e^\Phi+ \widetilde
\eta\Psi,}
one finds that the $[0,-{1\over 3}, -{2\over 3}, -1,-{4\over 3},-{5\over 3}]$
$C$-charge contribution to $(G+A)^2=0$ implies
\eqn\nonlt{Q A_0 =- A_0^2,\quad \{Q+A_0,A_{-{1\over 3}}
\}=0,\quad \{Q+A_0,A_{-{2\over 3}}\}=-A_{-{2\over 3}}^2,}
$$\widetilde
\eta A_0=-\{A_{-{2\over 3}}, A_{-{1\over 3}}\}, \quad \widetilde
\eta A_{-\th}
=-A_{-\tth}^2,\quad \widetilde
\eta A_{-\tth}=0.$$
The $[0,-{1\over 3},-{5\over 3}]$ $C$-charge equations are automatically
satisfied by \deft, while the 
$[-{2\over 3},-1,-{4\over 3}]$ $C$-charge equations imply the equations
of motion
\eqn\nonlint{
\widetilde
\eta(e^{-\Phi}(Q e^{\Phi}))= -\{\widetilde
\eta\Psi, e^{-\Phi}(Q\overline
\Psi)e^\Phi\},}
$$
Q(e^\Phi(\widetilde
\eta\Psi)e^{-\Phi})=-(Q\overline
\Psi)^2,\quad
\widetilde
\eta(e^{-\Phi}(Q\overline
\Psi) e^{\Phi})= -(\widetilde
\eta\Psi)^2.$$

Furthermore,
the equation  $(G+A)^2=0$ is invariant under the gauge transformation
$\d A = G\s +[A,\s]$ where
\eqn\defgs{\s = \s_{1\over 3} + \s_0 + \s_{-{1\over 3}}
= \{Q+A_0,\L_{1\over 3}\} + \widetilde
\eta\L_1 + \{A_{-\tth}
,\L_{2\over 3}\} +\widetilde
\eta\L_{2\over 3}.}
In terms of $[\Phi,\Psi,\overline
\Psi]$,
the gauge transformations are
\eqn\tunpan{\d e^\Phi = e^\Phi(\widetilde
\eta\L_1 + 
\{A_{-\tth},\L_{2\over 3}\} -\{A_{-\th}, \L_{1\over 3}\}) + (Q\L_0)e^\Phi,}
$$\d\Psi= \{Q+A_0,\L_{1\over 3}\} 
-\{A_{-\th},\L_{2\over 3}\} - 
\{A_{-\tth},\L_1\}  + \widetilde
\eta\L_{4\over 3},$$
$$
\d\overline
\Psi = e^\Phi(
\widetilde
\eta\L_{2\over 3} -\{A_{-\tth},\L_{1\over 3}\}) e^{-\Phi} +\{Q\overline
\Psi,\L_0\}
+Q\L_{-\th}.$$
So the linearized contribution to \nonlint\ and \tunpan\ reproduces
\tlin.

Note that if $[\Phi,\Psi,\overline
\Psi]$ did not have $C$-charges $[0,{1\over 3},
-{1\over 3}]$, the equations implied by $(G+A)^2=0$ would be inconsistent.
Firstly, $\Phi$ must have vanishing $C$-charge for $e^\Phi$ to have
well-defined $C$-charge. And secondly, for 
$(Q\overline
\Psi)^2$ 
and $(\widetilde
\eta\Psi)^2$ 
to have
the same $C$-charge as 
$Q(e^\Phi(\widetilde
\eta\Psi) e^{-\Phi})$ and
$\widetilde
\eta (e^{-\Phi} (Q\overline
\Psi) e^\Phi)$ in \nonlint,
$\Psi$ and $\overline
\Psi$
must have $C$-charge ${1\over 3}$ and $-\th$.
Also note that with more than three string fields, $(G+A)^2=0$
would imply inconsistent equations of motion. For example, with four
string fields, $(G+A)^2=0$ would imply equations with $C$-charges
$[0,-{1\over 4},-\half,-{3\over 4}, -1,-{5\over 4}, -{3\over 2},-{7\over 4}]$.
As in \nonlt, three of these equations could be satisfied by suitably
defining $A_n$. However, this would leave five equations of motion for
the four string fields.

\subsec{Open superstring field theory action}

To obtain the equations of motion of \nonlint\ from varying
$[\Phi,\Psi,\overline
\Psi]$ in an action, the $[-1,-\tth,-{4\over 3}]$
$C$-charge of these equations plus the $[0,-\th,\th]$ $C$-charge
of the string fields must equal the background $C$-charge. In other
words, the nonvanishing norm $\langle \xi e^{-2\phi}c\p c\p^2 c\rangle$
must carry $-1$ $C$-charge. With this assumption, one can easily check
that \nonlint\ comes from varying the action
\eqn\actone{S = \langle (e^{-\Phi}\widetilde
\eta e^\Phi)(e^{-\Phi} Q e^\Phi)
+\int_0^1 dt (e^{-\widehat\Phi}\p_t e^{\widehat\Phi})\{e^{-\widehat\Phi}
\widetilde
\eta 
e^{\widehat\Phi} , e^{-\widehat\Phi} Q e^{\widehat\Phi} \} }
$$+e^{-\Phi}(Q\overline
\Psi)e^\Phi (\widetilde
\eta\Psi)-\th\overline
\Psi (Q\overline
\Psi)^2 +
\th\Psi(\widetilde
\eta\Psi)^2\rangle$$
where $\widehat\Phi(t)$ is a function defined for $0\leq t\leq 1$
which satisfies $\widehat \Phi(0)=0$ and $\widehat \Phi(1)=\Phi$.

The first line of \actone\ is the same WZW-like action constructed in
\review\ for the NS sector. To give an interpretation for the second
line of \actone, note that the small Hilbert space norm
\eqn\small{\langle \widetilde\eta(\xi e^{-2\phi} c\p c \p^2 c)\rangle=
\langle  e^{-2\phi} c\p c \p^2 c\rangle= 1}
can be used when all fields in the correlation function are annihilated
by the $\eta$ zero mode \fms. Likewise, one can define a different
small Hilbert space norm
\eqn\smalltwo{\langle Q(\xi e^{-2\phi} c\p c \p^2 c)\rangle=
\langle  2\eta c\p c \rangle= 1}
when all fields in the correlation function are annihilated by $Q$.

These two small Hilbert spaces resemble chiral and antichiral superspaces
where the norm of \small\ is used for chiral $F$-terms and the norm of
\smalltwo\ is used for antichiral $F$-terms. So it is natural to define
a ``chiral'' field $\O$ as any field satisfying $\widetilde
\eta\O=0$, and an 
``antichiral'' field $\overline
\O$ as any field satisfying $Q\overline
\O=0$. To
distinguish the different Hilbert space norms, the notation $\langle \rangle_F$
and $\langle \rangle_{\overline
 F}$ will denote the small Hilbert space norms
of \small\ and \smalltwo\ respectively, 
and the notation $\langle \rangle_D$ will
denote the large Hilbert space norm.

Since $\langle \Psi(\widetilde
\eta\Psi)^2\rangle_D=
\langle (\widetilde
\eta\Psi)^3\rangle_F$ 
and $\langle \overline
\Psi(Q\overline
\Psi)^2\rangle_D=
\langle (Q\overline
\Psi)^3\rangle_{\overline
 F}$, the second line of \actone\
can be written as
\eqn\secondline{\langle e^{-\Phi}\overline
\O e^\Phi \O\rangle_D 
-\th\langle \overline
\O^3\rangle_{\overline
 F}
+\th\langle \O^3\rangle_{F}}
where $\O=\widetilde
\eta\Psi$ is a chiral string field and
$\overline
\O=Q\overline
\Psi$ is an antichiral string field. 
So the second line of \actone\ can be interpreted as 
the standard kinetic term and Yukawa potential for chiral
and antichiral fields. 
Note that 
the cohomologies of
$\widetilde
\eta $ and $Q$ are trivial in the large Hilbert space, so
any chiral superfield
$\O$ can be written as $\widetilde
\eta\Psi$ for some $\Psi$ and any antichiral superfield
$\overline
\O$ can be written as $Q\overline
\Psi$ for some $\overline
\Psi$.
One can therefore treat 
$\O$ and $\overline
 \O$ as fundamental chiral and antichiral string
fields in the action and forget about $\Psi$ and $\overline
\Psi$. Since $\langle \O^3\rangle_F$ cannot be written as a $D$-term
without introducing $\Psi$, one expects for the usual reasons that
this $F$-term does not receive quantum corrections.

\subsec{$Q$ with nonzero $C$-charge}

Although the construction of \actone\ assumed that $Q$
carries zero $C$-charge, this assumption can be slightly relaxed.
To preserve
the structure of the equations implied by $(G+A)^2=0$, it will be necessary
to assume only that $G=G_0 + G_{-{1\over 3}} + G_{-{2\over 3}}+ G_{-1}$. In
other words, it will be assumed that $Q$ and $\widetilde
\eta$ only contain terms
carrying $C$-charge $[0,-{1\over 3},-{2\over 3},-1]$. 
Note that $(Q+\widetilde
\eta)^2=0$ implies that $G_0^2=G_{-1}^2=0$, and it will
also be assumed that $G_0$ and $G_{-1}$ have trivial cohomology in
the large Hilbert space. With this assumption, a 
chiral string field $\O$ and antichiral
string field $\overline
\O$ can be defined by $G_{-1}\O=0$ and $G_0\overline
\O=0$, which implies that $\O= G_{-1}\Psi$
and $\overline\O=G_0 \overline\Psi$ for some $\Psi$ and $\overline\Psi$.

One can check that $(G+A)^2=0$ and $\d A= G\s +[A,\s]$ imply
consistent equations of motion and gauge invariances where
\eqn\ga{A_0 = e^{-\Phi}(G_0 e^\Phi) ,\quad A_{-\th}=
e^{-\Phi}(G_{-\th} e^\Phi) + e^{-\Phi}\overline
\O e^\Phi ,\quad A_{-\tth}=\O,}
$$\s_\th = 
\{G_0+A_0, \L_\th\}, \quad \s_0= G_{-1}\L_1 +\{G_{-\tth}+\O,\L_\tth\},
\quad \s_{-\th} = G_{-1}\L_\tth.$$
Defining $\cG_n = G_n + A_n$,
the equations of motion and
gauge invariances of \nonlint\ and \tunpan\ generalize to
\eqn\nleqg{ \{G_{-1},\cG_{0} \} = -\{\cG_{-\tth},\cG_{-\th}\},\quad
\{\cG_0, \cG_{-\tth}\} = -(\cG_{-\th})^2,\quad
\{G_{-1}, \cG_{-\th}\} = -(\cG_{-\tth})^2,}
\eqn\gaugeg{
\d e^\Phi = e^\Phi(G_{-1}\L_1 + \cG_{-\tth}\L_{\tth} -\cG_{-\th}\L_\th)
+(G_0\L_0) e^\Phi,}
$$\d\Psi = \cG_0\L_\th -\cG_{-\tth}\L_1 -\cG_{-\th}\L_\tth
+G_{-1}\L_{4\over 3},$$
$$\d\overline
\Psi = e^\Phi(G_{-1}\L_\tth-\cG_{-\tth}\L_\th)e^{-\Phi}
+\{G_{-\th}+\overline
\O,\L_0\} + G_0\L_{-\th}.$$

The action which produces the equations of motion of \nleqg\ is \sft :
\eqn\acttwo{
S = \langle (e^{-\Phi}G_{-1} e^\Phi)(e^{-\Phi} G_0 e^\Phi)
+(e^{-\Phi}G_{-\tth} e^\Phi)(e^{-\Phi} G_{-\th} e^\Phi)}
$$
+\int_0^1 dt (e^{-\widehat\Phi}\p_t e^{\widehat\Phi})(
\{e^{-\widehat\Phi}G_{-1}
e^{\widehat\Phi} , e^{-\widehat\Phi} G_0 e^{\widehat\Phi} \} 
+\{e^{-\widehat\Phi}G_{-\tth}
e^{\widehat\Phi} , e^{-\widehat\Phi} G_{-\th} e^{\widehat\Phi} \} )$$
$$+e^{-\Phi}\overline
\O e^\Phi \O
+\overline
\O e^\Phi (G_{-\tth} e^{-\Phi}) 
-\O e^{-\Phi} (G_{-\th} e^{\Phi}) \rangle_D$$
$$-\langle \half \overline
\O G_{-\th}\overline
\O + \th\overline
\O^3\rangle_{\overline
 F}
+
\langle \half \O G_{-\tth}\O + \th\O^3\rangle_{ F}$$
where $\langle \rangle_F$ and $\langle \rangle_{\overline
F}$ are defined
using the Hilbert space norms 
$\langle G_{-1}(\xi e^{-2\phi} c\p c\p^2 c)\rangle_F=1$ and
$\langle G_{0}(\xi e^{-2\phi} c\p c\p^2 c)\rangle_{\overline
 F}=1$.
Note that the chiral and antichiral $F$-terms in the last line of \acttwo\
rememble holomorphic and antiholomorphic Chern-Simons terms and are
not expected to receive quantum corrections.

There are two possible definitions of hermiticity which are consistent
with the action of \acttwo. The first possibility is that all string
fields $[\Phi,\Psi,\overline\Psi]$ and operators $G_n$ are independently
hermitian. The second possibility is that $\Phi$ is antihermitian,
$\overline\Psi$ is the hermitian conjugate of $\Psi$, and
$G_n$ is the hermitian conjugate of $G_{-1-n}$.\foot{Using the second
hermiticity definition, the action of \acttwo\ naively appears to
be imaginary. However, in the explicit example considered
in subsection (4.3), $\xi e^{-2\phi}
c \p c\p^2 c$ will be imaginary with this definition. So if one defines
$\langle \xi e^{-2\phi} c \p c\p^2 c\rangle =1$, the action of
\acttwo\ is real since the norm is imaginary.}

\newsec{Splitting the States into Three String Fields}

\subsec{Conditions for the $C$-charge}

In this section, the action of \acttwo\ will be made explicit
by giving two examples of $C$-charge.
As discussed in
section 3,
consistency of the action implies that
the $C$-charge must be a conserved charge
with the following properties:
1) All superstring states must carry $C$-charge 0 or $\pm\th$;
2) All terms in $Q+\widetilde
\eta$ must carry $C$-charge $[0,-\th, -\tth, -1]$ where the terms
with $0$ and $-1$ $C$-charge have trivial cohomology;  
and
3) The large Hilbert space background charge $\xi e^{-2\phi} c\p c\p^2 c$
must carry $C$-charge $-1$.

Since the term $\eta\p\eta e^{2\phi} b$ in $Q$ is the term with
trivial cohomology, this term should carry zero $C$-charge. And since
both
$\eta$ and $\xi e^{-2\phi} c\p c\p^2 c$ must carry $C$-charge $-1$,
$e^{n\phi}$ must carry $C$-charge $n$ and $(b,c)$ must carry $C$-charge
zero. This implies that $C=P+\th N$ where $P$ is picture and $N$
is some conserved charge constructed from the RNS matter fields.
Furthermore, since $3C$ must be an integer, $N$ must be chosen such
that NS states carry integer $N$-charge and R states carry half-integer
$N$-charge.

In a flat background, examples of such $N$-charges are
\eqn\Ncharge{N = \sum_{j=1}^J \int \psi^{2j-2}\psi^{2j-1}}
for $J=1$, $J=3$ or $J=5$. These examples (up to Wick rotations)
manifestly preserve an SO($10-2J$)$\times$ U($J$) subgroup of
the d=10 Lorentz group. The examples $J=1$ and $J=3$ will be
explicitly discussed below, and the example $J=5$ can be treated
similarly if one ignores hermiticity questions.

Note that $J$ must be odd in \Ncharge\ in order that R states
carry half-integer $N$-charge. This dependence on $J$ might seem strange,
but at the end of section 5, an R-R kinetic term will be constructed with
d=$10-2J$ Lorentz invariance. When $J$ is even, the Type IIB R-R
spectrum in d=$10-2J$ contains a self-dual $(5-J)$-form field
strength, so one expects to find problems with constructing
an action.

\subsec{Manifest d=8 Lorentz covariance}

Splitting the Ramond states into different string fields implies
that the sixteen component d=10 spinor must split into two eight
component spinors. So the maximum Lorentz subgroup which can be manifestly
preserved is d=8 Lorentz covariance. As will now be shown, this can
be achieved by defining
\eqn\ceight{C=P +\th\int \psi^0\psi^9}
where $P$ is the picture and $\int \psi^0\psi^9$ is the 
SO(1,1) charge in the $M_{09}$ direction. An SO(1,1) boost direction
has been chosen, so following the discussion at the end of section 3,
one can use the first hermiticity definition in which
$\Psi$ and $\overline\Psi$ are independent hermitian
string fields. If one had instead chosen a U(1) rotation direction
(e.g. $C=P +{i\over 3}\int \psi^1\psi^2$), one would use the second
hermiticity definition in which $\Psi$ and $\overline\Psi$ are
hermitian conjugate string fields.

With the choice of \ceight, $Q+\widetilde
\eta$ splits into terms of $C$-charge
$[0,\th,-\th,-1]$ where the terms of $C$-charge $\pm\th$ are
$\eta e^\phi \p x^- \psi^+$ and
$\eta e^\phi \p x^+ \psi^-$ using the notation
$x^\pm = {1\over \sqrt{2}}(x^0 \pm x^9)$
and $\psi^\pm =
{1\over \sqrt{2}}(\psi^0 \pm \psi^9)$.
To remove the unwanted term of $C$-charge $+\th$,
one can perform the similarity transformation
\eqn\similt{Q+\widetilde
\eta \to e^{R}(Q+\widetilde
\eta) e^{-R} \quad{\rm where}\quad
R= \int c\xi e^{-\phi}\psi^+\p x^-.} To show that \similt\
only has terms with $C$-charge $[0,-\th,-\tth,-1]$, it is convenient
to use the result of
\ref\chan{J.N. Acosta, N. Berkovits and O. Chand\'{\i}a, 
{\it A Note on the Superstring BRST Operator}, Phys. Lett. B454 (1999) 247,
hep-th/9902178.} where the RNS BRST operator was 
written as $Q= e^{-S} (-\int\eta\p\eta e^{2\phi} b) e^S$
with
\eqn\simils{S=\int 
(c\xi e^{-\phi}\psi^m \p x_m +\half \p\phi c\p c\xi\p\xi e^{-2\phi})
\quad{\rm and ~~}m=0 {\rm~~ to~~} 9.}
So the similarity transformation of \similt\ takes $Q+\widetilde
\eta$ into
\eqn\compu{
e^{R}(Q+\widetilde
\eta) e^{-R} = e^{R} e^{-S} (-\int\eta\p\eta e^{2\phi}b)
 e^S e^{-R} + e^{R}\widetilde \eta e^{-R}}
$$= e^{-U} (-\int\eta\p\eta e^{2\phi} b) e^U 
+\int\eta e^\phi \psi^-\p x^+
+\int c e^{-\phi} \psi^+ \p x^- + \widetilde\eta$$ 
\eqn\similt{{\rm where}\quad U=\int 
(c\xi e^{-\phi}\psi^k \p x_k +\half (\p\phi+\psi^0\psi^9)
c\p c\xi\p\xi e^{-2\phi})\quad
{\rm and ~~}k=1 {\rm~~ to~~} 8.}

So after performing the similarity transformation of \similt,
$G= e^{R}(Q+\widetilde\eta) e^{-R}$ 
only contains terms of $C$-charge $[0,-\th,-\tth,-1]$
which are given by
\eqn\gggdef{G_0 =  e^{-U} (-\int \eta\p\eta e^{2\phi}b) e^U,\quad
G_{-\th} = \int \eta e^\phi\psi^- \p x^+,}
$$
G_{-\tth} =  \int c e^{-\phi}\psi^+ \p x^-,\quad G_{-1}=\int \eta.$$
With $G_n$ defined by \gggdef\ 
and the string fields $[\Phi,\Psi,\overline\Psi]$
defined using \ceight, 
\acttwo\ gives a manifestly d=8 Lorentz covariant open
superstring field theory action. 

Note that \ceight\ implies that the massless gluon $A_m(x)$ and
gluino $\chi^\a(x)$ split into the following components of the string fields:
\eqn\spliteight{\Phi = \xi c e^{-\phi} \psi^k A_k(x) + ... , 
\quad
\Psi = \xi c e^{-\phi} \psi^+ A_+(x) + \xi c e^{-\half\phi}\Sigma_a\chi^a(x)
+ ... ,}
$$ \overline
\Psi = \xi c e^{-\phi} \psi^- A_-(x) + \xi\p\xi c \p c 
e^{-{5\over 2}\phi}\Sigma_{\dot a}\overline\chi^{\dot a}(x) + ...,$$
where $\chi^a = (\g^+\chi)^a$ and
$\overline\chi^{\dot a} = (\g^-\chi)^{\dot a}$ are the SO(8) components
of $\chi^\a$. One can check that the gluino contribution 
\eqn\rcont{S= \half 
Tr\int d^{10} x  ~ \chi^\a \g^m_{\a\b}(\p_m \chi^\b + [A_m,\chi^\b] )}
$$
= Tr\int d^{10} x (\chi^a \s^k_{a\dot b}(\p_k \overline\chi^{\dot b}
+ [A_k,\overline\chi^{\dot b}])
+\half\overline\chi^{\dot a}(\p_-\overline\chi^{\dot a}
+ [A_-,\overline\chi^{\dot a}])
+\half\chi^a (\p_+ \chi^a + [A_+,\chi^a]) )$$
comes from the terms
\eqn\comeseight{\langle e^{-\Phi}\overline\O e^\Phi\O\rangle_D
- \langle \half \overline\O G_{-\th}\overline\O + 
\th\overline\O^3\rangle_{\overline F}
+ \langle \half \O G_{-\tth}\O + \th\O^3\rangle_F}
in \acttwo.

\subsec{Manifest N=1 d=4 super-Poincar\'e covariance}

A second possible choice for the $C$-charge is
\eqn\cfour{C=P +{i\over 3}\int (\psi^4\psi^5 +\psi^6\psi^7+\psi^8\psi^9).}
This charge easily generalizes to $C=P+\th\int \p H$ for compactification
on a Calabi-Yau threefold with U(1) current $J_{CY}=\p H$. Because $\p H$
is antihermitian,
the string fields and operators must satisfy
the second hermiticity definition, i.e. 
\eqn\hermfour{\Phi^\dagger =-\Phi,\quad \Psi^\dagger=\overline\Psi,\quad
G_0^\dagger = G_{-1},\quad G_{-\th}^\dagger= G_{-\tth}.}
As will now be shown,
the above hermiticity conditions are natural if one rewrites the RNS
worldsheet variables in terms of d=4 Green-Schwarz-like variables \ref\covcy
{N. Berkovits, {\it Covariant Quantization of the Green-Schwarz    
Superstring in a Calabi-Yau Background}, Nucl. Phys. B431 (1994) 258,
hep-th/9404162.}, which also allows 
N=1 d=4 super-Poincar\'e covariance to be made manifest. 

The first step to constructing an N=1 d=4
super-Poincar\'e covariant action is to perform the similarity
transformation
\eqn\simfour{Q+\widetilde
\eta \to e^{R+\half U}(Q+\widetilde
\eta) e^{-R -\half U} \quad{\rm where}}
$$U=\int 
(c\xi e^{-\phi}\psi^p \p x_p +\half (\p\phi+\p H)
c\p c\xi\p\xi e^{-2\phi}),\quad
R= \int c\xi e^{-\phi}\psi^{+j}\p x^{-j},$$
$p=0$ to 3, $j=1$ to 3, 
$\psi^{\pm j}= {1\over \sqrt 2}(\psi^{2j+2}\pm i\psi^{2j+3})$ and
$x^{\pm j}= {1\over \sqrt 2}(x^{2j+2}\pm i x^{2j+3})$.
Since $Q=e^{-S} (-\int\eta\p\eta e^{2\phi} b) e^S$
where $S$ is defined in \simils, one finds
\eqn\resfour{e^{R+\half U}(Q+\widetilde
\eta) e^{-R -\half U} =}
$$ 
 e^{-\half U} (-\int\eta\p\eta e^{2\phi} b) e^{\half U }
+\int\eta e^\phi \psi^{-j}\p x^{+j}
+\int c e^{-\phi} \psi^{+j} \p x^{-j} + e^{\half U}\widetilde\eta e^{-\half U}
.$$
Equation \resfour\ can be written in manifestly N=1 d=4 super-Poincar\'e
covariant notation by defining the d=4 Green-Schwarz-like variables \covcy
\eqn\hybrid{\t^\a = e^{\half\phi}\Sigma^\a e^{-\half H},\quad \overline\t^\ad= 
c\xi e^{-{3\over 2}\phi}\Sigma^\ad e^{\half H},\quad
p_\a = e^{-\half\phi}\Sigma_\a e^{\half H},\quad \overline p_\ad= 
b\eta e^{{3\over 2}\phi}\Sigma_\ad e^{-\half H},}
$$\p\rho = 3\phi + c b + 2\xi \eta -\p H, \quad \Gamma^{+j}=\xi 
e^{-\phi}\psi^{+j},
\quad \Gamma^{-j}=  \eta e^\phi \psi^{-j},$$
where $\Sigma^\a$ and $\Sigma^\ad$ are d=4 spin fields constructed from
$[\psi^0,\psi^1,\psi^2,\psi^3]$ and $(\a,\ad)=1$ to 2. 
Note that the variables of \hybrid\
are GSO-projected and satisfy free-field OPE's.

In terms of these variables, one can check that 
\eqn\ressusy{e^{R+\half U}(Q+\widetilde
\eta) e^{-R -\half U} =}
$$ \int [\half d^\a d_\a e^\rho + \Gamma^{-j}\p x^{+j} +
\half\epsilon^{jkl} \p x^{-j}\Gamma^{-k}\Gamma^{-l} e^{-\rho} +
{1\over {12}}\overline d^\ad \overline d_\ad e^{-2\rho}
\epsilon^{jkl} \Gamma^{-j}\Gamma^{-k}\Gamma^{-l}] ,$$
where 
$d_\a=p_\a +{i\over 2}
\tb^\ad \p x_p 
\sigma^p_{\a\ad}-{1\over 4}(\tb)^2 \p\t_\a +{1\over {8}}\t_\a\p(\tb)^2$
and
$\overline d_\ad=\overline p_\ad + {i\over 2}
\t^\a \p x_p \sigma^p_{\a\ad}-{1\over 4}(\t)^2
\p\tb_\ad +{1\over {8}}\tb_\ad\p(\t)^2$
are supersymmetric combinations of the fermionic momenta.
Since $C= {1\over 3}\int \Gamma^{+j}\Gamma^{-j}$, 
\ressusy\ implies that
\eqn\defops{
G_0 = \int \half d^\a d_\a e^\rho,\quad G_{-\th}= \int \Gamma^{-j}\p x^{+j}, }
$$
G_{-\tth}=\int
\half\epsilon^{jkl} \p x^{-j}\Gamma^{-k}\Gamma^{-l} e^{-\rho},\quad 
G_{-1}=\int {1\over {12}}\overline d^\ad \overline d_\ad e^{-2\rho}
\epsilon^{jkl} \Gamma^{-j}\Gamma^{-k}\Gamma^{-l} .$$
The hermiticity properties of \hermfour\ are satisfied if one
defines 
\eqn\satf{\t_\a^\dagger=\tb_\ad,\quad p_\a^\dagger=-\overline p_\ad, \quad
(\Gamma^{-j})^\dagger = 
\half\epsilon^{jkl} \Gamma^{-k}\Gamma^{-l} e^{-\rho} ,}
$$
(\Gamma^{+j})^\dagger = 
\half\epsilon^{jkl} \Gamma^{+k}\Gamma^{+l} e^{\rho} ,\quad
(e^\rho)^\dagger = {1\over 6}
e^{-2\rho}\epsilon^{jkl} \Gamma^{-j}\Gamma^{-k}\Gamma^{-l} .$$
Note that $\xi e^{-2\phi} c\p c \p^2 c= {1\over {24}}(\t)^2 (\tb)^2
e^{-\rho}\epsilon^{jkl} \Gamma^{-j}\Gamma^{-k}\Gamma^{-l}$
is imaginary, as discussed in footnote 5. Since $j_{ghost}=
\p\rho +\Gamma^{-j}\Gamma^{+j}$, the ghost-number zero string fields
$[\Phi,\Psi,\overline\Psi]$ carry $\rho$-charge $[0,+1,-1]$.

So
using the operators of \defops\ in the action of \acttwo, one gets
a manifestly N=1 d=4 super-Poincar\'e covariant open superstring field
theory action. As shown in \sft, the massless contribution to this action
reproduces the d=10 super-Yang-Mills action written in terms of 
N=1 d=4 superfields \ref\marc{N. Marcus, A. Sagnotti and W. Siegel,
{\it Ten-Dimensional Supersymmetric Yang-Mills Theory
in Terms of Four-Dimensional Superfields}, Nucl. Phys. B224 (1983) 159.}
where the $D$-terms and $F$-terms in \acttwo\ reproduce the standard
N=1 d=4 superspace $D$-terms and $F$-terms.

\newsec{Open Questions}

In this paper, it was argued that construction of a consistent
open superstring field theory action requires splitting the superstring
states into three string fields
which carry conserved $C$-charge 0 and $\pm\th$. Different choices
for splitting the superstring states produce different actions with
different manifest symmetries. This construction raises several
obvious questions.

One question is how to generalize the action of \acttwo\ to include
the GSO($-$) superstring states which are present for non-BPS
$D$-branes and for $D$-brane/anti-$D$-brane configurations \ref\sens
{A. Sen, {\it Tachyon Condensation on the Brane Antibrane System},
JHEP 9808 (1998) 010, hep-th/9805019.}.
As will be shown in a separate paper \ref\bps{
N. Berkovits and C.T. Echevarria,
to appear.} with Carlos Tello Echevarria, these
GSO($-$) states can be easily included by adjoining $2\times 2$ 
matrices to the three string fields and operators of \acttwo, as was done
for the NS action in \ref\openns{N. Berkovits, {\it The Tachyon Potential
in Open Neveu-Schwarz String Field Theory}, JHEP 0004 (2000) 022,
hep-th/0001084\semi N. Berkovits, A. Sen and B. Zwiebach, 
{\it Tachyon Condensation in Superstring Field Theory}, Nucl. Phys.
B587 (2000) 147, hep-th/0002211.}. Such an action might be useful
for studying broken supersymmetry before tachyon condensation as
proposed by Yoneya \ref\yon{T. Yoneya, {\it Worldsheet String
Duality and Hidden Supersymmetry}, hep-th/0109058\semi
T. Yoneya, {\it Spontaneously Broken Spacetime Supersymmetry
in Open String Theory without GSO Projection}, Nucl. Phys. B576 (2000)
219, hep-th/9912255.}.

A second question is if the different actions produced by different
splittings are related by a field redefinition. It is easy to show
that the equations of motion of \nleqg\ for different splittings are related
by a field redefinition since \nleqg\ can be written in a splitting-independent
manner as $(G+A)^2=0$. So any solution to the equations of motion using
one splitting is also a solution using another splitting. However, it is
not obvious that there exists an off-shell field redefinition which
relates the different actions.

A third question is which worldsheet conformal field theory backgrounds
allow construction of an open superstring field theory action, i.e. which
backgrounds allow definition of a $C$-charge with the desired properties.
It might seem strange that not all N=1 c=15 superconformal field theory
(scft) backgrounds allow construction of an open superstring field theory
action. However, this should not be too surprising since, for example,
R-R backgrounds cannot be described by an N=1 c=15 scft since they
mix the RNS matter and ghost fields. 

To describe R-R backgrounds \ref\wv{N. Berkovits, C. Vafa and
E. Witten, {\it Conformal Field Theory of AdS Background with
Ramond-Ramond Flux}, JHEP 9903 (1999) 018, hep-th/9902098\semi
N. Berkovits, {\it Quantization of the Superstring in Ramond-Ramond
Backgrounds}, Class. Quant. Grav. 17 (2000) 971, hep-th/9910251.}
\ref\bzz{N. Berkovits, M. Bershadsky, T. Hauer, S. Zhukov and
B. Zwiebach, {\it Superstring Theory on $AdS_2\times S^2$ as
a Coset Supermanifold}, Nucl. Phys. B567 (2000) 61, hep-th/9907200.},
one needs to embed the superstring in a worldsheet N=2 c=6 scft
\topo. An open superstring field theory action in a worldsheet
N=2 scft background can be defined by replacing $G=Q+\widetilde\eta$
of \acttwo\ with $G=\int (G^+ + \widetilde G^+)$ where $G^+$
and $\widetilde G^+$ are constructed from the fermionic worldsheet
N=2 generators as explained in \topo. For example, the action 
constructed in subsection (4.3) generalizes to an $AdS_2\times S^2$
background with R-R flux by replacing the flat d=4 Minkowski background
with the N=2 scft described in \bzz. It would be interesting to
know precisely which N=2 c=6 scft backgrounds allow construction of
an open superstring field theory action.

A final question is if the methods of this paper are useful for
constructing a closed superstring field theory action. Although
the NS-NS contribution to such an action can be constructed as
in \ref\pcs{R. Saroja and A. Sen, {\it Picture Changing Operators
in Closed Fermionic String Field Theory}, Phys. Lett. B286 (1992)
256, hep-th/9202087.}, the only successful construction up to
now of a kinetic term for the R-R sector
\ref\sun{N. Berkovits, {\it Manifest Electromagnetic Duality in
Closed 
Superstring Field Theory}, Phys. Lett. B388 (1996) 743, hep-th/9607070.}
uses
the SU(1,1)
formalism \ref
\SZ{W. Siegel, {\it Covariantly Second Quantized String II},
Phys. Lett. B151 (1985) 391\semi W. Siegel and B. Zwiebach,
{\it Gauge String Fields}, Nucl. Phys. B263 (1986) 105.} of Siegel
and Zwiebach.
However, even for bosonic string field theory, the SU(1,1)
formalism has not yet been generalized to include interactions.
Furthermore, the R-R kinetic term of \sun\
involves an infinite number of fields,
which is not surprising because of the self-dual five-form in the
Type IIB R-R sector.

Since the closed string field can be understood as the ``left-right''
product of two open string fields, the methods of this paper
suggest introducing a closed
superstring field $\Phi_{m,\widehat n}$ carrying left-moving $C$-charge
$m$ for $m\in [0,\pm\th]$ and right-moving $C$-charge $\widehat n$ for
$\widehat n
\in [0,\pm\th]$. When $Q$ carries zero $C$-charge, one can construct
the closed superstring kinetic term 
\eqn\cskone{S_{closed}=
\sum_{m,\widehat n=-\th}^{\th} \langle\Phi_{-m,-\widehat n} 
(c-\widehat c)_0
(Q+\widehat Q)\widetilde\eta\widetilde{\widehat\eta}~
\Phi_{m,\widehat n}\rangle,}
which is the natural closed string generalization of the open
superstring kinetic term $\langle \Phi Q \widetilde\eta\Phi\rangle$. Unlike
the kinetic term of \rrk, \cskone\ is gauge invariant when
$\Phi_{m,\widehat n}$ satisfies the constraint 
$(b-\widehat b)_0\Phi_{m,\widehat n}=0$. The kinetic term of \cskone\
can be written in a more symmetric form as
\eqn\csktwo{S_{closed}=
\sum_{m,\widehat n=-\th}^{\th} \langle\Phi_{-m,-\widehat n} 
(c\eta+\widehat c\widehat \eta)_0
(Q+\widehat Q)(\eta+\widehat\eta)_0
~\Phi_{m,\widehat n}\rangle}
since the $(b-\widehat b)_0$ constraint implies that only
the $(c-\widehat c)_0 (\eta-\widehat\eta)_0$ part of 
$(c\eta+\widehat c\widehat \eta)_0$ contributes to the action.
Note that \csktwo\ can be generalized to any N=(2,2) c=6 scft
as
\eqn\cskthr{ S_{closed}=
\sum_{m,\widehat n=-\th}^{\th} \langle\Phi_{-m,-\widehat n} 
(J^{++}+\widehat J^{++})_0
(G^+ +\widehat G^+)_0
(\widetilde G^+ +\widehat{\widetilde G}{}^+)_0 ~\Phi_{m,\widehat n}\rangle}
where $(G^+,\widetilde G^+,J^{++})$ 
and $(\widehat G^+,\widehat{\widetilde G}{}^+,\widehat J^{++})$ 
are constructed from the left and right-moving N=2
superconformal generators as described in \topo.

Unfortunately, the kinetic term of \cskone\ does not seem to generalize
when $Q$ carries non-zero $C$-charge, i.e. when $G=Q+\widetilde\eta=
G_0 + G_{-\th} + G_{-\tth} + G_{-1}$. 
Nevertheless, using the $C$-charges constructed in section 4, 
one could consider defining \cskone\ where $Q$ and $\widetilde\eta$
are replaced by $G_0$ and $G_{-1}$. Using the $C$-charge of subsection
(4.2), this would give the eight-dimensional contribution to the
kinetic term, i.e. the contribution from string fields which are
independent of $x^\pm$ and $\psi^\pm$. And using the 
the $C$-charge of subsection
(4.3), this would give the four-dimensional contribution to the
kinetic term, i.e. the contribution from string fields which are
independent of $x^{\pm j}$ and $\Gamma^{\pm j}$. Note that these
four and eight-dimensional contributions do not contain self-dual
field strengths in the R-R sector, so one does not expect any problems.
However, if one could have constructed a $C$-charge which preserved
d=$10-2J$ Lorentz invariance for $J$ even, one would expect problems
since there are self-dual $(5-J)$-form field strengths in the
$(10-2J)$-dimensional contribution to the R-R kinetic term.

\vskip 15pt
{\bf Acknowledgements:} 
I would like to thank Carlos Tello Echevarria, Cumrun Vafa and
Barton Zwiebach   
for useful discussions, and
CNPq grant 300256/94-9, 
Pronex grant 66.2002/1998-9,
and FAPESP grant 99/12763-0
for partial financial support.
This research was partially conducted during the period the author was
employed by the Clay Mathematics Institute as a CMI Prize Fellow.

\listrefs

\end